\def\comment#1{}
\def\old#1{}
\def\mn#1{*\marginpar[*#1]{*#1}}
\def\mn#1{}
\documentstyle[12pt,epsf]{article}

 \def\lfrac#1#2{{{{#1}/{#2}}}}
\begin{document}
\begin{center}
\vskip 0.001in

\hfill Journal ref.: Phys. Rev. {\bf B59} 12083 (1999)

\vskip1cm

{\Large \bf Nonperturbative $XY$-model approach to
strong coupling superconductivity
in two and three dimensions.}

\vskip 2cm

  { \bf  Egor Babaev}
\footnote{On leave from
A.F.Ioffe Physico-Technical Institute
Russian Academy of Sciences,
Politechnicheskaja str. 26,
St. Petersburg, 194021, Russia.~~~~~
Email:~ egor@teorfys.uu.se}

\vskip0.3cm
{  \it
 \ Institut f\"ur Theoretische Physik \\
Freie Universit\"at Berlin, Arnimallee 14, 1000 Berlin 33, Germany}

{  \it and \ Institute for Theoretical Physics, Uppsala University \\
    Box 803, S-75108 Uppsala, Sweden}

\vskip1cm

{\bf Hagen Kleinert}
\footnote{ Email:~
 kleinert@physik.fu-berlin.de ~ URL:
http://www.physik.fu-berlin.de/\~{}kleinert ~~ Phone/Fax:
 0049 30 8383034.}

\vskip0.3cm
{ \it  \ Institut f\"ur Theoretische Physik \\
Freie Universit\"at Berlin, Arnimallee 14, 1000 Berlin 33, Germany}

\vskip0.5cm

 \vskip 0.15in

\end{center}
\begin{abstract}
For
 an electron gas with
 $ \delta $-function attraction we investigate the
crossover from weak- to strong-coupling supercoductivity
in two and three dimensions.
We derive analytic expressions for
the stiffness of phase fluctuations
and set up effective XY-models
which serve to determine nonperturbatively
 the temperature of phase decoherence
where superconductivity breaks down.
We find the transition temperature
$T_c$ as a monotonous function of the coupling strength and carrier density both in two
and three dimensions, and give analytic  formulas for the
merging of the temperature of phase
decoherence with the temperature of pair formation in the
weak-coupling limit.

\end{abstract}
\baselineskip 0.6cm


\newpage

\section{Introduction}

The crossover from BCS superconductors  to a
 Bose-Einstein condensate of tightly bound
fermion pairs was first  studied
many years ago in
Refs.~\cite{Ea}--\cite{Le}
in a model with $\delta$-function attraction.
This crossover has recently raised renewed interest \cite{Uem}- \cite{ek},
especially after the doping dependence of the critical temperature ($T_c$)
and non-Fermi-liquid behavior  above $T_c$ were combined in a phenomenological
discussion of High-$T_c$ cuprates \cite{@Uem}, \cite{Em}.
In this paper we present a detailed study of
the crossover based on a nonperturbative
procedure in which derivative expansion
is used to set up an effective XY-model whose
well-known
nonperturbative properties
render information
on the entire crossover regime of
the above model.

Physically, the most important distinctions
between conventional (BCS) and strong-coupling (Bose-Einstein)
regime
lies in the fact that in the former
only a small fraction of the
conduction electrons
is paired
with
the superfluid density involving all pairs, whereas
in the latter
practically all  carriers are paired
below a certain temperature $T^*$, although not condensed.
The temperature
has to be lowered further below some critical temperature
$T_c < T^*$ to make
these pairs condensed and
establish
 phase coherence, which leads to
superconductive behavior.

In the model to be investigated in this paper,
the
crossover
from BCS-type to Bose-type
superconductivity
will take place
either by
varying the coupling
strength, or by
decreasing
the carrier density.

Since in the BCS theory pair coupling is weak, it can well
be described by mean-field theory for the pair fields.
In the opposite limit of strong pair binding, on the other hand,
superconductivity set in as a macroscopic
occupationis of ${\bf q=0}$ level and
we are obliged to go beyond mean field
level to describe such an effect.
In three dimensions, crossover from BCS
superconductivity to the Bose-Einstein condensation
of tightly bound fermion pairs
was first investigated  in Ref.~\cite{Noz}
by summing particle-particle ladder diagrams
which
correspond to
Gaussian fluctuations
around the mean field, in the functional integral formalism
it was studied in Ref.~\cite{R8}. In both
papers, fluctuation corrections
were retained in the number equation which
was solved together with mean-field
gap equation.
In these papers,
starting from fermionic system, gas of electron
pairs was maped in the strong-coupling limit
to the ideal Bose gas and critical temperature
was shown to becomes
asymptoticaly temperature of Bose-Einstein
condensation of ideal Bose gas of particles
with mass $2m$ and density $n/2$, where
 $m$ and $n$ are fermion mass and density. In
this aproximation $T_c$ has an arificial
maximum at intermediate coupling strength
thus approaching limiting value in the strong-coupling
limit from above. This artifact was removed
in the generalized self-consistent Green's function
numerical approach in \cite{h}.

In this paper we shall study the properties of collective modes
with help of the lowest gradient terms governing the
Gaussian fluctuations
around the mean-field solution. These are most violent
in the
phase of the order parameter.
Phase transitions
in a system with these
fluctuations are well understood in
two and three dimensions from extensive
studies of the XY-model.
By setting up an equivalent XY-model
we are therefore able to
describe well
the onset and disappearance of superconductivity
in the entire crossover regime.
In this way, we shall obtain
simple formulas for
the critical temperature $T_c$
as a monotonously increasing
function
of the
coupling strength and carrier density in
both two and three dimensions.
In the  weak-coupling  limit, we give simple explicit
formulas
which show how
the temperature of the
$XY$-model transition
converges to the transition
temperature in the
BCS theory.

The weak- to strong-coupling superconductivity crossover
in two dimensions
was studied via the Kosterlitz-Thouless theory
in Refs.~\cite{Dr}, \cite{shar}.
In Ref.~\cite{shar}
it was investigated within
the same
model as ours at a fixed carrier density,
 but only
numerically\footnote{We shall see in Section~4
that these numerical results do not cover
the entire
crossover region, in particular, the merging of $T_{KT}$
and $T^*$ in the weak-coupling region  is missing - the
effect that we show analyticaly in our paper.}.
The different properties of size and phase fluctuations
was also exploited in Ref. \cite{Trav}.

\section{Model}

The Hamiltonian of our model is the typical BCS Hamiltonian
in $D$ dimensions ($\hbar=1$)
\begin{eqnarray}
H &=& \sum_\sigma \int \! d^D x
        \, \psi_\sigma^{\dag} ({\bf x})
        \left(-{{\bf \nabla}^2 \over 2m} -\mu\right)
        \psi_\sigma({\bf x})
         + g \int\!d^D x\,
        \psi_\uparrow^{\dag}({\bf x}) \psi_\downarrow^{\dag}({\bf x})
        \psi_\downarrow^{\phantom{\dag}}({\bf x})
        \psi_\uparrow^{\phantom{\dag}}({\bf x})
        \label{1.0},
\end{eqnarray}
where $\psi_\sigma({\bf x})$ is the Fermi field operator,
$\sigma=\uparrow,\downarrow$
denotes the spin components,
$m$ is the effective fermionic mass, and
$g < 0 $  the strength of an attractive potential
$ g  \delta ({\bf x} - {\bf x}')$.

The mean-field equations for the
gap parameter $\Delta$ and the chemical potential $\mu$
are obtained  in the standard way
 (see for example \cite{6'}):

\begin{eqnarray}
-{1\over g} &=& \frac{1}{V} \sum_{\bf k} {1\over 2 E_{\bf k}}
\tanh{E_{\bf k} \over 2T} ,\label{1.1}\\
  n &=& \frac{N}{V} = {1\over V} \sum_{\bf k} \left(1-{\xi_{\bf k}
 \over E_{\bf k}} \tanh{E_{\bf k} \over 2T}\right),
\label{1.2}
\end{eqnarray}
where the sum runs over all
wave vectors
$\bf k$,
 $N$ is the total number
of fermions,
 $V $  the
volume of the system,
 and
\begin{equation}
 E_{\bf k}=\sqrt{\xi_{\bf k}^2 + \Delta^2}
{}~~~\mbox{with}~~~
\xi_{\bf k} = {{\bf k}^2 \over 2 m} - \mu
\label{1.3}
\end{equation}
 are the energies of single-particle excitations.

\old{
Changing the  sum over {\bf k} to  an integral
over $\xi$
and over the directions of ${\bf k}$, on which the integrand does not depend,
 we arrive
in three dimensions at the gap equation:
\begin{equation}
{1 \over g} = \kappa_{3} \int_{-\mu}^\infty d\xi {\sqrt{\xi+\mu}
\over 2 \sqrt{\xi^2+\Delta^2}} \tanh{\sqrt{\xi^2 +\Delta^2} \over 2 T}
\label{1.3.1},
\end{equation}
where the constant $\kappa_{3}= m^{3/2} / \sqrt{2} \pi^2$ has dimension
energy$^{-3/2}$/volume.
In two-dimensions, the  density of
states is constant, and the gap equation becomes
\begin{equation}
{1 \over g} = \kappa_{2} \int_{-\mu}^\infty d\xi  { 1
\over 2 \sqrt{\xi^2+\Delta^2}} \tanh{\sqrt{\xi^2 +\Delta^2} \over 2 T}
\label{1.3.2},
\end{equation}
with a constant $ \kappa_{2}= m/2\pi$  of dimension energy$^{-1}$/two-volume.
In two dimensions, the particle number in Eq.~(\ref{1.2}) can
be integrated
with the result:
\begin{equation}
n=\frac{m}{2 \pi} \left\{
\sqrt{\mu^{2} + \Delta^{2}} + \mu +
2 T \log{\left[
1 + \exp{\left(-\frac{\sqrt{\mu^{2} + \Delta^{2}}}{T}\right)}
\right]}\right\},
\label{numb}
\end{equation}
the right-hand side being a function $n(\mu, T, \Delta) $.
}

The $\delta$-function potential produces an
artificial divergence and requires
 regularization. A BCS superconductor possesses
 a natural cutoff supplied by the Debye frequency $ \omega _D$.
For the crossover problem
to be treated here
this is no longer a useful quantity, since in the strong-coupling
limit all
 fermions
participate in the interaction, not  only those
in a thin shell of width $ \omega _D$ around the Fermi surface.
To be applicable in this regime,
 we  renormalize
the gap equation in three dimensions with the help of the
experimentally observable
$s$-wave scattering length $a_s$,
for which the  low-energy limit of the
two-body scattering process gives an equally divergent
expression \cite{h}--\cite{r2}:
\begin{equation}
        {m \over 4 \pi a_s}
=
{1\over g}
        + {1\over V}
\sum_{\bf k}
{m \over {\bf k}^2} .
 \label{1.4}
\end{equation}
Eliminating $g$ from (\ref{1.4}) and  (\ref{1.1})
we obtain a renormalized gap equation
\begin{equation}
        -{m \over 4 \pi a_s} = {1\over V} \sum_{\bf k}
        \left[{1\over 2 E_{\bf k}} \tanh{E_{\bf k} \over 2T}
 - {m \over {\bf k}^2} \right],
        \label{1.5}
\end{equation}
 in which $1/k_Fa_s$ plays the role
of a dimensionless coupling constant which monotonically increases
 from $-\infty$ to $\infty$ as the bare
coupling constant $|g|$ runs from small
(BCS limit) to  large values
(BE limit).
This equation is to be solved
simultaneously with (\ref{1.2}).
These mean-field equations were
first analyzed
at a fixed carrier density in Refs.~\cite{R8} and \cite{R-rev}.
Here we shall first
reproduce some of the obtained
estimates for $T^*$ and $\mu$.

In the BCS limit, the chemical potential $\mu $
does not differ much
from the Fermi energy
 $\epsilon_F$, whereas with increasing interaction strength,
  the distribution function $n_{\bf k}$
 broadens and $\mu $ decreases.
In the BE limit we have tightly bound
pairs and nondegenerate fermions with a large negative chemical
potential $|\mu|\gg T$.
In the strong coupling limit Eq.~ (\ref{1.5}) provide us
an estimation for $T^*$ -
characteristic temperature of the thermal pair breaking
\cite{R8},
whereas Eq.~ (\ref{1.2}) determines $\mu$. From
Eq~. (\ref{1.2}) we obtain that in the BE limit $\mu = - E_b/2$,
 where $E_b=1/m a_s^2$ is the binding energy of the bound pairs.
In the BE limit,  we can estimate that the pseudogap sets in at
$T^* \simeq E_b/2 \log(E_b / \epsilon_F)^{3/2}$.
A simple ``chemical'' equilibrium estimate
$(\mu_b=2\mu_f)$ yields  for the temperature
of pair
 dissociation: $T_{\rm dissoc} \simeq E_b/\log(E_b/\epsilon_F)^{3/2}$
which shows at strong couplings
$T^*$ is indeed related to  pair formation   \cite{R8},
\cite{RR8} (which in the strong-coupling regime
lies above  the temperature of the onset of
phase coherence  \cite{Noz}--\cite{ranrev}).

The gap in the spectrum
of  single-particle excitations has a special feature \cite{Le}, \cite{Ea},
\cite{R-rev}
when the chemical potential changes its sign.
The sign change occurs at the minimum of the Bogoliubov
quasiparticle energy $E_{\bf k}$ where this energy
defines the gap energy in the quasiparticle spectrum:
\begin{eqnarray}
E_{\rm gap}={\rm min}\left(\xi_{\bf k}^2 +\Delta^2 \right)^{1/2} .
\label{1.5.1}
\end{eqnarray}
Thus, for positive chemical potential,
the gap energy is given directly by the gap function $ \Delta$, whereas for
negative chemical potential,
it is larger than that:
\begin{eqnarray}
E_{\rm gap}=  \Biggl\{ \matrix{
\Delta & {\rm for}  \ \ \ \mu >0 ,\cr
(\mu^2 +\Delta^2)^{1/2} & {\rm for} \ \ \ \mu < 0 .\cr}
\label{1.5.2}
\end{eqnarray}

In two dimensions, a nonzero bound state energy $\epsilon_0$
exists for any coupling strength. The cutoff
can therefore be eliminated by
subtracting from the two-dimensional zero-temperature gap equation
\cite{pist}-\cite{2Danalit}
\begin{equation}
- \frac{1}{g} = \frac{1}{2V} \sum_{\bf k} \frac{1}{\sqrt{\xi_{\bf k}^2 +
\Delta^2}}=
\frac{m}{4 \pi} \int_{- x_0}^{\infty} d z \frac{1}{\sqrt{1+ z^2}},
\end{equation}
where $ z=k^2/2m\Delta-x_0$, $x_0= \mu / \Delta $
the bound-state equation
\begin{equation}
-\frac{1}{g}=\frac{1}{V}\sum_{\bf k} \frac{1}{ {\bf k}^2/m + \epsilon_0}=
\frac{m}{2\pi}\int_{-x_0}^\infty d z \frac{1}{ 2z+\epsilon_0/\Delta+2x_0}.
\label{@binding}\end{equation}
After performing the elementary integrals,
one finds:
\begin{equation}
\frac{\epsilon_0}{\Delta}=\sqrt{1+x_0^2} - x_0.
\label{@gal2D}\end{equation}
\old{
{}From Eq.~(\ref{numb}) we see that at zero temperature,
gap and chemical potential
are related to $x_0$ by
\begin{eqnarray}
   {\Delta \over \epsilon_F }
   &=&
   {2 \over x_0 +\sqrt{1+x_0^2}},
   \label{1.13}
\end{eqnarray}
\begin{eqnarray}
   {\mu \over \epsilon_F}
   &=&
   {2 x_0 \over x_0 +\sqrt{1+x_0^2}}.
   \label{1.14}
\end{eqnarray}
The two relations are plotted in Fig.~\ref{f2}.
Combining (\ref{@gal2D}) with (\ref{1.13})
we find
the dependence of the ratio $ \lfrac{\epsilon_0}{\epsilon_F} $
on the crossover parameter $x_0$:
\begin{equation}
\frac{\epsilon_0}{\epsilon_F}=
2 \frac{\sqrt{1+x_0^2} -x_0}{\sqrt{1+x_0^2} +x_0 }
\label{q1}
\end{equation}
\begin{figure}[tbh]
\input 2d0.tps
\caption{
Gap function $\Delta$
and chemical
potential $\mu$ at zero temperature
as functions of $x_0$
in two dimensions.
}
\label{f2}\end{figure}
}
In the next sectioins
we will work at finite tempetrature,
in doing so, we
do not fix the carrier density
but assume
the presence of a reservoir which provides us with a
temperature-independent chemical potential
$\mu=\mu( 1/k_Fa_s ; T=0 )$.
Such a fixed $\mu$
will be most convenient
for deriving simple analytic
results for the finite-temperature behavior of the system.
In this fixed $\mu$ model carrier
density becomes temperature dependent.\footnote{In Ref.~\cite{shar},
the temperature dependence of the
chemical potential
was calculated numerically for the entire
crossover region
within a "fixed carrier density model", where it turned out to be very small
in comparison with the dependence on the coupling strength.}
For experimental measurements of $\mu$ see \cite{Rie}

\section[Phase Fluctuations and Kosterlitz-Thouless
Transition]
{Phase fluctuations in Two Dimensions \\ and Kosterlitz-Thouless Transition}

In this section we make use of derivative
expansion to determine
the relevant
stiffness parameter
for the study of
phase fluctuations, which in two dimensions
determines the temperature of
 the Kosterlitz-Thouless transition.
In
two-dimensional system,
the phase fluctuations
are most violent causing the
strongest
modifications of
the mean-field properties.
The
 Coleman-Mermin-Wagner-Hohenberg theorem
 \cite{col}
forbids the
existence of a strict long-range order,
but there is quasi-long-range order manifesting itself
in a
power behavior of
the correlation functions at all temperatures below $T_{KT}$.
We shall neglect the coupling to the
magnetic vector potential throughout the upcoming
discussion, so that
the phase coherence below
$T_c$ can be of long or quasi-long range, unspoiled by the Meissner effect
which would reduce the range to
a finite
penetration depth.

The effective Hamiltonian
from which we deduce the stiffness of the phase fluctuations
was derived in \cite{shak}, \cite{shar}.
In this Section we summarize a few important aspects of it,
with a reminder of its derivation given below.
Writing the spacetime-dependent order parameter
as $ \Delta(x)e^{i\theta({x})}$,
where $x$ denotes the four-vector $x=(\tau ,{\bf x})$
formed from imaginary time and position vector,
the partition function
may be written as a functional integral
\cite{shak,W,T}
\begin{equation} Z(\mu, T) = \int \Delta \,{\cal D} \Delta\,
{\cal D} \theta \exp{[-\beta \Omega (\mu, T, \Delta(x), \partial \theta
(x))]},
\label{bk1}
\end{equation}
where
\begin{equation}
\beta \Omega ( \mu , T, \Delta (x), \partial \theta (x)) =
\frac{1}{g} \int_{0}^{\beta}
d \tau \int d {\bf x} \Delta^{2}(x) -
\mbox{Tr log} G^{-1} + \mbox{Tr log} G_{0}^{-1}
\label{dfg}
\end{equation}
is  the one-loop effective action,
containing the
inverse Green function of the fermions
in the collective pair field
\begin{eqnarray}
G^{-1} & = & -\hat{I} \partial_{\tau} +
\tau_{3} \left(\frac{\nabla^{2}}{2 m} + \mu \right) +
\tau_{1} \Delta(\tau, \mbox{\bf x})
\nonumber \\
& - &
\tau_{3} \left[ \frac {i \partial_{\tau} \theta(\tau, \mbox{\bf x})}{2} +
\frac{(\nabla \theta(\tau, \mbox{\bf x}))^{2}}{8m} \right] +
\hat{I} \left[\frac{i \nabla^{2} \theta(\tau, \mbox{\bf x})}{4 m} +
\frac{i \nabla \theta(\tau, \mbox{\bf x}) \nabla}{2m} \right].
\label{bk3}
\end{eqnarray}
Here $\tau_1 ,~ \tau_3$ are the usual Pauli matrices,
and
$G_{0} = G|_{\mu, \Delta, \theta=0}$ is added for
regularization.

Let us now assume that
phase gradients are small. Then
$\Omega(\mu, T, \Delta(x), \partial \theta(x))$
can be approximated as follows:
\begin{equation}
\Omega (\mu, \Delta(x), \partial \theta(x))  \simeq
\Omega _{\rm kin} (\mu, T, \Delta, \partial \theta(x)) +
\Omega _{\rm pot} (\mu, T, \Delta),
\label{bk4}
\end{equation}
with the ``kinetic" term (see \cite{shak})
\begin{equation}
\Omega _{\rm kin} (\mu, T, \Delta, \partial \theta(x))
 =  T \mbox{Tr} \sum_{n=1}^{\infty}
\left. \frac{1}{n} ({\cal G} \Sigma)^{n} \right|_{\Delta = \rm const}
\label{bkt5}
\end{equation}
and
the ``potential" term
\begin{equation}
\Omega _{\rm pot} (\mu, T, \Delta)  =
\left. \left(\frac{1}{g} \int d^D x \Delta^{2} -
T \mbox{Tr log} {\cal G}^{-1} +
T \mbox{Tr log} G_{0}^{-1} \right) \right|_{\Delta = \rm const}.
\label{bkt5a}
\end{equation}
The latter coincides with the
mean-field energy,
 determining
the modulus of $\Delta (\mu , T)$
and thus
the stiffness of phase fluctuations.
The kinetic part  $\Omega_{\rm kin}$ contains gradient
terms whose size is determined by
the modulus of $\Delta (\mu , T)$.
Given the stiffness,
one may immediately set up an equivalent
XY-model.
 Both $\Omega_{\rm kin}$ and $\Omega_{\rm pot}$
are expressed in terms of the Green function of the fermions,
which solves the equation
\begin{equation}
\left[-\hat I \partial_{\tau} +
\tau_{3} \left(\frac{\nabla^{2}}{2 m} + \mu \right)
+ \tau_{1} \Delta \right]
{\cal G}(\tau, \mbox{\bf x}) = \delta(\tau) \delta(\mbox{\bf x})
\label{bkt6}
\end{equation}
and
\begin{equation}
\Sigma(\partial \theta) \equiv
\tau_{3} \left[ \frac {i \partial_{\tau} \theta}{2}  +
\frac{(\nabla \theta)^{2}}{8 m} \right] -
\hat{I} \left[\frac{i \nabla^{2} \theta}{4 m} +
\frac{i \nabla \theta(\tau, \mbox{\bf x}) \nabla}{2m} \right].
\label{bkt7}
\end{equation}
The gradient expansion that we use to determine stiffness
 was first made in Ref. ~\cite{shak} at zero temperature.
In Ref.~\cite{shar}, the kinetic term $\Omega_{\rm kin}$ was
calculated in two dimensions at finite temperature for arbitrary chemical
potential  retaining terms with $n= 1,2$
in the expansion (\ref{bkt5}).

 The result is
\begin{equation}
\Omega _{\rm kin} =
\frac{T}{2}
\int_{0}^{\beta} d \tau \int d^D x
\left[ n(\mu, T, \Delta ) i \partial_\tau \theta +
J(\mu, T, \Delta(\mu, T)) (\nabla \theta)^{2} +
K(\mu, T, \Delta(\mu, T)) (\partial_{\tau} \theta)^{2}
\right],
\label{bk8}
\end{equation}
where $J(\mu, T, \Delta)$ is the stiffness coefficient whose explicit form is
\begin{equation}
J(\mu, T, \Delta) = \frac{1}{4m} n(\mu, T, \Delta) -
\frac{T}{4\pi} \int_{-\mu/2T}^{\infty} dx
\frac{x + \mu/2T}{\cosh^{2} \sqrt{x^{2} +
{ \Delta^{2}}/{ 4 T^{2}}}},
\label{bk9}
\end{equation}
The other coefficents are:
\begin{equation}
K(\mu, T, \Delta) = \frac{m}{8 \pi} \left(
1 + \frac{\mu}{\sqrt{\mu^2 + \Delta^2}}
\tanh{\frac{\sqrt{\mu^2 + \Delta^2}}{2T}} \right),
\label{bk10}
\end{equation}
and   $n(\mu, T, \Delta)$
is the  density of fermions (\ref{1.2})
which varies with temperature in our model.
At the temperature $T^*$ where
the modulus of $\Delta$ vanishes, also the stiffness
disappears.

We are now ready to specify the effective
XY-model
governing the phase fluctuations.
The model Hamiltonian corresponding to the
gradient term in $\theta(x)$
is
\cite{bkt}, \cite{GFCM}:
\begin{eqnarray}
H={J \over 2} \int d {\bf x} [\nabla \theta({ \bf x})]^2.
\end{eqnarray}
In contrast to the
standard XY-model,
the stiffness parameter is not a constant
but depends
on temperature via the solution of
gap and number
equations (\ref{1.1}) and (\ref{1.2}).
Clearly,  Kosterlitz-Thouless transition always take
place below $T^*$.
For vortices of a high fugacity,
the temperature
of the phase transition is determined by
the well-known formula \cite{Min}
\begin{eqnarray}
T_{\rm KT}={\pi \over 2}J  \label{KT},
\label{@jinT}\end{eqnarray}
which follows from the
divergence of the average square size
of a vortex-antivortex pair.
Since these attract each other by a Coulomb potential
$v(r)=2\pi J\log(r/r_0)$,
the average square distance is
\begin{equation}
<r^2>\propto \int_{r_0}^\infty dr r \,r^2 e^{-(2\pi J/T)\log (r/r_0)}
\propto \frac{1}{4-2\pi J/ T},
\label{@}\end{equation}
which diverges indeed at the temperature (\ref{KT}).
In our case $T_{KT}$ should be determined self-consistently:
\begin{equation}
T_{\rm KT}=\frac{\pi}{2} J(\mu, T_{\rm KT}, \Delta(\mu, T_{\rm KT})).
\label{e1}
\end{equation}
{}From (\ref{bk9}),  (\ref{1.2}) and     (\ref{KT})
it is easily seen that $T_{\rm KT}$ indeed tends to zero when
the pair attraction vanishes in which case $\Delta (T=0) = 0$.
In general, the behavior
 of $T_{\rm KT}$
for strong and weak couplings
is found by the following considerations.
We observe
that the particle number $n$
does not vary appreciably in these limits
with temperature in the range $0 < T < T^*$,
so that weak-coupling
estimates for $T_{\rm KT}$ derived
within the model with temperature-independent chemical potential
(i.e., when the system is coupled to a large reservoir
) practically
coincide with those derived from a
fixed fermion density.
Further it is
immediately realized
that
in the weak-coupling limit
$\Delta (T_{\rm KT},\mu)/T_{\rm KT}$ is a small
parameter.
At zero coupling, the stiffness $J(\mu,T_{\rm KT},\Delta (\mu, T_{\rm KT}))$
vanishes identically, such that  an estimate
of $J$ at  weak couplings
requires
calculating a lowest-order
correction
to the second term of eq.(\ref{bk9})
 proportional to
 $\Delta(T_{\rm KT}, \mu)/T_{\rm KT} $.
Thus weak-coupling expression for
stiffness reads:
\begin{equation}
J(T) \simeq \frac{7 \zeta (3)}{ 16 \pi^3} \epsilon_F
\frac{\Delta^2(T)}{T^*{}^2}.
\end{equation}
Equating this with  the stiffness in
(\ref{@jinT}) we obtain
 the weak-coupling equation for $T_{\rm KT}$:
\begin{equation}
T_{\rm KT} \simeq
\frac{ \epsilon_F}{4}
\left( 1- \frac{T_{\rm KT}}{T^*} \right).
\label{e2}
\end{equation}
where  $\epsilon_F =(\pi / m) n$
is
the
Fermi energy of free fermions.
It is useful to introduce reduced dimensionless temperatures
${\tilde T_{KT}} \equiv T_{KT} / \epsilon_F$ and
${\tilde T^*} = T^* / \epsilon_F$
which are small in the weak-coupling limit.
Then we rewrite Eq.~(\ref{e2}) as
\begin{equation}
{\tilde T_{KT}} \simeq
\frac{1}{4}
\frac{1}{1+1 / 4{\tilde T^*}}.
\end{equation}
For small ${ \tilde T}^*$ we may expand
\begin{equation}
{\tilde T_{KT}} \approx {\tilde T^*} - 4 {\tilde T^*}{}^2.
\label{s01}
\end{equation}
This equation shows
 explicitly how for
decreasing coupling strength
$T_{\rm KT}$
merges with $T^*$.

\begin{figure}[tbh]
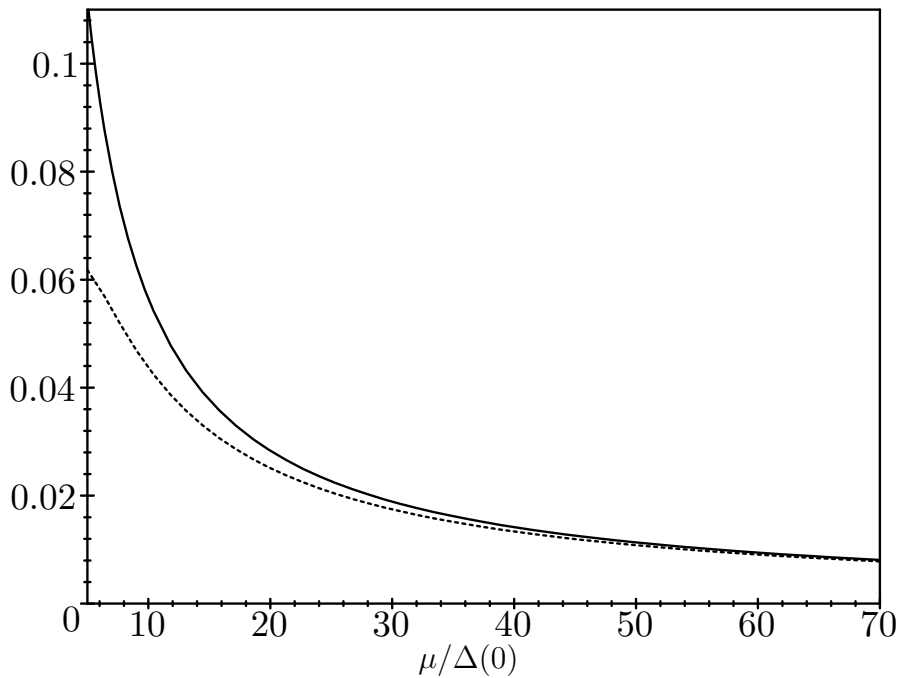

\input 2d.tps  \\[-2cm]
\caption{
Weak-coupling behavior of $T_{KT}$.
The solid line is $T^*$, the
dashed line represent $T_{KT}$.}
\label{f7}\end{figure}

For  weak coupling strengths,
$T_{\rm KT}$ behaves like
\begin{eqnarray}
{T}_{\rm KT} \approx
\frac{e^\gamma}{\pi} {\Delta(0)}.
\label{c1}
\end{eqnarray}
The merging of the two temperatures
in the  weak-coupling regime is displayed
in Fig.~\ref{f7}.

Consider now the opposite limit of strong couplings.
There we see from Eqs. (\ref{e1}),
(\ref{1.1})
(\ref{1.2}), and
(\ref{bk9})
for $T_{\rm KT}$, $n(T, \mu)$, and $\Delta (T, \mu)$
that
$T_{\rm KT}$ tends to a constant value.
We can observe that in the strong-coupling limit
$\Delta(T_{\rm KT})$ is always
situated close to
the zero-temperature value of
$\Delta (T_{\rm KT}, \mu) \approx \Delta (T=0,\mu )$.
Taking this into the account we derive
an estimate for the second term in (\ref{bk9}),
thus obtaining the strong-coupling equation for $T_{\rm KT}$:
\begin{eqnarray}
T_{\rm KT}
\simeq \frac{\pi}{8} \left\{
\frac{1}{m}n -
\frac{T_{\rm KT}}{\pi} \exp\left[ -
 \frac{\sqrt{\mu^2+\Delta^2(T_{\rm KT},\mu)}}{T_{\rm KT}}
\right]
\right\}
\label{e30}
\end{eqnarray}
%
%
With the
approximation
$\Delta (T_{\rm KT}, \mu) \approx \Delta (T= 0, \mu)$
we find that the first term in the exponent tends
in the strong-coupling limit to a constant,
$ \Delta^2 (T_{\rm KT}, \mu)/2 \mu T_{\rm KT} \rightarrow - 4$,
whereas the first  term in the brackets
tends to $- \infty$,
so that Eq.~(\ref{e30})
has the limiting form
\begin{eqnarray}
T_{\rm KT} \simeq
\frac{\pi}{8} \frac{n}{m}
\left\{
1 - \frac{1}{8} \exp\left[
\frac{2\mu}{\epsilon_F} -4
\right]
\right\}.
\label{e32}
\end{eqnarray}
%
 Thus  for increasing coupling strength,
the phase-decoherence temperature $T_{\rm KT}$
tends very quickly towards a constant:
\begin{equation}
T_{KT} \simeq  \frac{\pi}{8} \frac{n}{m}.
\label{e302}
\end{equation}
In this limit
we know from Eq.~(\ref{1.2}) that the difference
in the carrier density at zero temperature, $n(T=0)$,
becomes equal to $n(T=T_{KT})$,
so that our
limiting result coincides
with that obtained in
the "fixed carrier density model":
\begin{equation}
T_{\rm KT} = \frac{\epsilon_F(n_0)}{8} = \frac{\pi}{8m} n_0,
\label{e4}
\end{equation}
where we have inserted $\epsilon_F(n)=(\pi/m)n$
for the
Fermi energy of free fermions
at the carrier density
$n_0=n(T=0)$.

From the above
asymtotical formulas
for weak- and strong-coupling limits
we see
that the temperature of the Kosterlitz-Thouless
transition
is
a monotonous function of coupling
strength and carrier density.
The crossover takes place
in a very narrow region  $\lfrac{\mu}{\Delta (0)} \in(-1,1)$.
It is also observed in
the behavior
of three-dimensional
condensation temperature $T_c$  of the
gas of tightly bound, almost free composite bosons.
In Refs.~ \cite{Noz}, \cite{R8}
which include only quadratic fluctuations around the mean field
(corresponding to ladder diagrams),
$T_c$ was shown to tend
to a constant free Bose gas value
 $T_c=[n/2 \zeta(3/2)]^{2/3} \pi /m$,
with no dependence on
the internal
structure of the boson.

Here we find a similar result in two dimensions,
where $T_{KT}$ tends to a constant depending only on
the mass $2m$ and the density $n/2$ of the pairs.
No dependence on the coupling strength is left.
The only difference with respect to the three-dimensional case
is that here the
transition temperature $T_c=T_{KT}$
is
linear in the carrier density $n$, while growing like $n^{2/3}$
in three dimensions.
Our limiting result (\ref{e4}) agrees
with
Ref.~\cite{Dr} and ~\cite{shar}.
There exists a
corresponding equation
for the temperature $T^*$
in the strong-coupling
limit
$\epsilon_0 \gg \epsilon_F$:
\begin{equation}
T^* \simeq \frac{\epsilon_0}{2}\frac{1}{ \log\lfrac{\epsilon_0}{\epsilon_F}}.
\end{equation}

\old{
For experimental purposes,
the dependence
of the ratio
$2 \Delta(0) / T_{KT}$
on the coupling strength
is of interest. It is plotted
  in Fig.~\ref{f8}.
Analytically, we have
in the weak-coupling
limit
\begin{equation}
\frac{2 \Delta(0)}{T_{KT}} =
\frac{2 \pi}{e^\gamma}
\left\{
1+\frac{e^\gamma}{\pi}\frac{4}{x_0} +
\left[
\frac{1}{8} +
\left(
\frac{4 e^{\gamma}}{\pi}
\right)^2
\right]\frac{1}{x_0^2}
\right\} +{\cal O}\left(x_0^{-3}\right)  ,
\end{equation}
and for strong-couplings:
\begin{equation}
\frac{2\Delta(0)}{T_{KT}} \simeq
\frac{32}{\sqrt{x_0^2+1}+x_0} \simeq -64 x_0
{}.
\end{equation}
The two curves can easily be interpolated graphically
to all
coupling strengths, as seen in the figure.
\begin{figure}[tbh]
\input 8.tps  \\[-2cm]
\caption{
Weak-coupling  and strong-coupling
estimates for the ratio
 $2\Delta (0)/T_{KT}$ (solid curves).
The dashed line is a graphical interpolation.
}
\label{f8}\end{figure}
}

\section[]
{Phase Fluctuations in Three Dimensions
}

 In this section we discuss in a completely analogous
way the fluctuations in three dimensions, where
stiffness is, for small temperatures where $\Delta(T)$
is close to $\Delta(0)$:
\begin{equation}
J_{\rm 3D}(\mu,T, \Delta)
=\frac{1}{4m}n(\mu,T,\Delta)-
\frac{ \sqrt{2m}}{16 \pi^2}\frac{1}{T}
\int^\infty_{-\mu}
d \xi
\frac{(\xi+\mu)^{3/2}}{\cosh^2(\sqrt{\xi^2+\Delta^2}/2T)},
\label{@finitev}\end{equation}
governing the phase fluctuations via an effective XY-model
\begin{equation}
H= \frac{ J_{\rm 3D}}{2} \int d^3 { \bf x} [ \nabla \theta( {\bf x})]^2.
\end{equation}
The temperature of the phase transition in this model can
reasonably be estimated using mean-field methods for the lattice 3D XY-Model
\cite{GFCM}:
\begin{equation}
T_{\rm 3D}^{MF} \simeq 3 J_{\rm 3D}a,
\end{equation}
 $a=1/n_b^{1/3}$ is the lattice spacing of the theory \cite{GFCM}
where $n_b$ is number of pairs.

\begin{figure}[tbh]
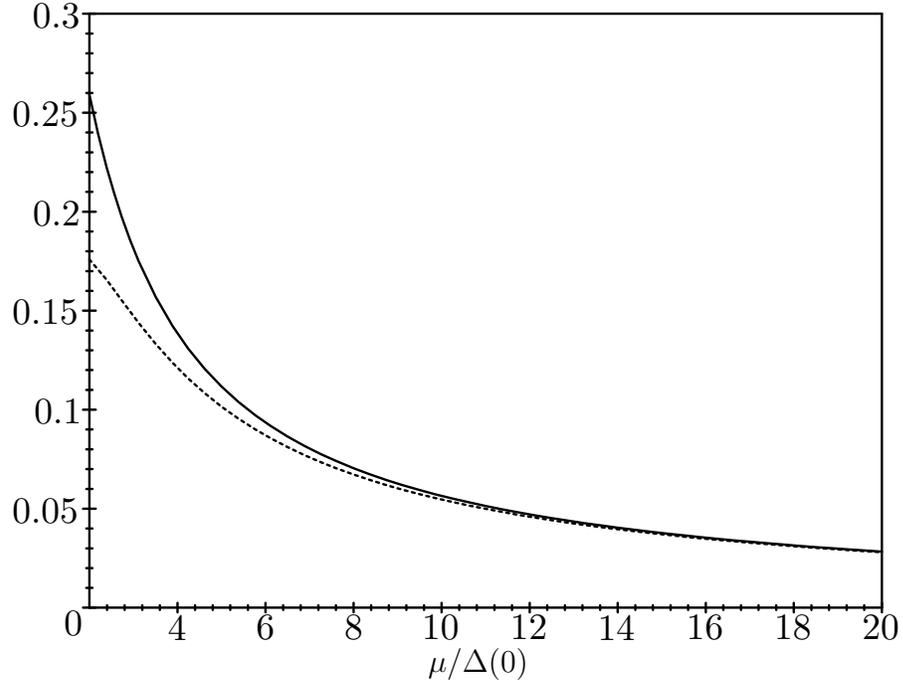

\input 3d.tps  \\[-2cm]
\caption{
Weak-coupling behavior of $T_c$
in three dimesions.
The solid line is $T^*$,
$T_c$ is plotted with dashed line.
}
\label{3d}\end{figure}

In the weak-coupling limit, the stiffness coefficent
may be derived with the help of Gorkov's well-known
method (setting $ T_c \approx T^*$):
\begin{equation}
J_{\rm 3D}=\frac{7}{48 \pi^4} \zeta(3) \frac{p_F^3}{m} \frac{\Delta^2}{T^*{}^2},
\label{@stiff}\end{equation}
This  is precisely
the coefficient
of the gradient term in the Ginzburg-Landau expansion,
In the weak-coupling limit the two temperatures merge
according to the formula:
\begin{equation}
{\tilde T_c }= {\tilde T^* } - \alpha {\tilde T^*{}^{5/2}},
\end{equation}
which contains a larger power
of ${\tilde T^*}$ in the
second term as well as
a smaller prefactor $\alpha=(2 \pi^2)^{2/3}/2\approx3.65$,
as
compared with the two-dimensional
separation formula
(\ref{s01}).
The merging behavior is displayed in Fig. (\ref{3d}).

In the strong-coupling limit of the theory
where we have tightly bound composite bosons,
the phase stiffness tends asymptotically to:
\begin{equation}
J=\frac{n}{4m} - \frac{3 \sqrt{2 \pi m}}{16 \pi^2} T^{3/2}
\exp\left[-\frac{\sqrt{\mu^2+\Delta^2}}{T}
\right],
\label{@stiff@}
\end{equation}
%
It obviously tends in this limit quickly to
\begin{equation}
J_{BE}=\frac{n}{4m}.
\end{equation}
An estimate for the critical temperature, obtained
via the
mean-field treatment of the 3D XY-model
on the
lattice reads in this limit:
\begin{equation}
T_c=
\frac{3}{2m} \left[ \left(\frac{n}{2}\right)^{2/3}-
\frac{1}{n^{1/3} } \frac{1}{2^{ 7/6}\pi^{3/2}} T_c^{3/2} m^{3/2}
\exp\left( -\frac{\sqrt{\mu^2+\Delta^2}}{T_c}
\right)
\right]
\label{chaoslabs}
\end{equation}
This quickly tends from below to the  value:
\begin{equation}
T_c^{ \rm 3D XY} = \frac{3 n^{2/3}}{2^{5/3} m}=
\epsilon_F \frac{3}{(6 \pi^2)^{2/3}} \simeq 0.2 \epsilon_F.
\end{equation}
This result is very close to
the temperature of the condensation of bosons
of mass $2m$
and density $n/2$, which, as it was discussed in the introduction
was obtained including
the effect of Gaussian fluctuations
into the mean-field equation for the particle number \cite{Noz,R8}
yielding\footnote{In the
study of the critical temperature
via gaussian corrections to the number
equation,
 mentioned in the introduction,
 the crossover behavior of the critical temperature
shows an artificial maximum in the region of intermediate couplings
\cite{Noz,R8}, so
that the
limiting value is approached from above in the strong-coupling limit.
This is in contrast to our three-dimensional $XY$-model approach. }
\begin{equation}
T_c^{\rm Bosons} =[n/2\zeta(3/2)]^{2/3} \pi/m =0.218 \epsilon_F.
\label{@bosons}
\end{equation}

\section{ Conclusion}

We have studied
the crossover
from BCS-type to Bose-type
superconductivity.
For this purpose
we have used the gradient expansion
of the effective energy functional
to set up an equivalent XY-model
which allows us to investigate
the onset of long-range order in the phase fluctuations.
In two dimensions, we have given
a simple analytic expression
which shows
how
the resulting
Kosterlitz-Thouless temperature
$T_{\rm KT}$
at which quasi-long-range order sets in
 moves towards the pair-binding temperature
$T^*$, and merges with it in the weak-coupling limit.
A similar expression was found
in three dimensions.
In the strong coupling limit
we find that critical temperature tends
in both two and three
dimensions
to a constant value
as the chemical potential
changes it sign.

Let us finally remark that the separation of $T^*$
and $T_c$ has an  analogy
in the ferroelectrics and magnets
which also contain two
separate characteristic temperatures, for example in the latter case---
the Stoner- and the Curie-temperature.
It also can be studied more precisely in a simple field theoretic model
in 2+$ \epsilon $ dimensions with an O($n$)
symmetry for large $n$. In such a model,
the existence of two small paramters $ \epsilon $ and $1/n$ has permitted
us recently to
{\em prove\/} the existence of two transitions,
and to exhibit clearly their different physical origins
\cite{GN}.

\section{Acknowledgments}
We thank Profs. K. Bennemann,  K. Maki, and
 V.Emery  for
explaining to us some aspects of $T^*$ crossover in superconductive cuprates
and to Proffessor Y. Uemura for useful points.
One of us (E.B.) is grateful to all members of Prof.~Kleinert's
group at the  Institut f\"ur Theoretische Physik
Freie Universit\"at Berlin for their kind
hospitality, and Dr. S.G. Sharapov
for useful discussions.
~\\                ~\\
{\bf Note added in proof}:   \\
Here and in our earlier preprint cond-mat/9804206
we criticize the discussion of the weak-coupling
behavior of superconductive phase transition
in two dimensions given
in Ref.~\cite{shar}.
However, after our paper went to print in
Phys. Rev. B {\bf 59}, 12083 (1999)),
the authors of \cite{shar} updated their discussion in
their preprint cond-mat/9709034 version 2, which
appeared
in JETP {\bf 88}, 685 (1999), and
our criticism no longer applies.

\vskip 0.6cm
\vskip 0.4cm

\end{document}